*Review*

# QoS Challenges and Opportunities in Wireless Sensor/Actuator Networks

Feng Xia [1,2]

1   Faculty of Information Technology, Queensland University of Technology,
    Brisbane QLD 4001, Australia
2   College of Computer Science and Technology, Zhejiang University, Hangzhou 310027, China

E-mail: f.xia@ieee.org; f.xia@acm.org

**Abstract:** A wireless sensor/actuator network (WSAN) is a group of sensors and actuators that are geographically distributed and interconnected by wireless networks. Sensors gather information about the state of physical world. Actuators react to this information by performing appropriate actions. WSANs thus enable cyber systems to monitor and manipulate the behavior of the physical world. WSANs are growing at a tremendous pace, just like the exploding evolution of Internet. Supporting quality of service (QoS) will be of critical importance for pervasive WSANs that serve as the network infrastructure of diverse applications. To spark new research and development interests in this field, this paper examines and discusses the requirements, critical challenges, and open research issues on QoS management in WSANs. A brief overview of recent progress is given.

## 1. Introduction

Wireless sensor networks (WSNs) [1,2] have been studied for about 10 years. Today this field is widely supported by an increasing number of dedicated journals such as ACM Trans. on Sensor Networks and Int. J. of Distributed Sensor Networks, conferences such as SENSYS (ACM Conf. on Embedded Networked Sensor Systems), IPSN (ACM/IEEE Int. Conf. on Information Processing in

Sensor Networks), and DCOSS (IEEE Int. Conf. on Distributed Computing in Sensor Systems), and commercial companies such as Crossbow, Ember, Sentilla, Dust Networks, Microsoft, Intel, and Sun Microsystems, to mention just a few. Numerous special issues of renowned journals on sensor networks have been published and special sessions of leading conferences organized, with more expected to appear in the future. Since its inception, WSN has grown into a hot research area at a tremendous pace. A large number of institutions and researchers around the world have set their feet in this field, and launched various research and development projects. Significant advances have been achieved in almost all aspects, including architecture, hardware, software, system design, supporting tools, standards, applications, etc [3].

WSNs are designed to gather information about the state of physical world and transmit sensed data to interested users, typically used in applications like habitat monitoring, military surveillance, agriculture and environmental sensing, and health monitoring. In most cases, they are unable to effect on the physical environment. In many applications, however, only observing the state of the physical system is not sufficient; it is also expected to respond to the sensed events/data by performing corresponding actions upon the system. For instance, in a fire handling system, the actuators need to turn on the water sprinklers upon receipt of a report of fire. This need for actuation heralds the emergence of wireless sensor/actuator networks (WSANs) [4-7], a substantial extension of sensor networks that feature coexistence of sensors and actuators. WSANs enable the application systems to sense, interact, and change the physical world, e.g., to monitor and manipulate the temperature and lighting in a smart office or the speed and direction of a mobile robot. It is envisioned that WSANs will be one of the most critical technologies for building the network infrastructure of future cyber-physical systems [8]. They will revolutionize the way we interact with the physical world.

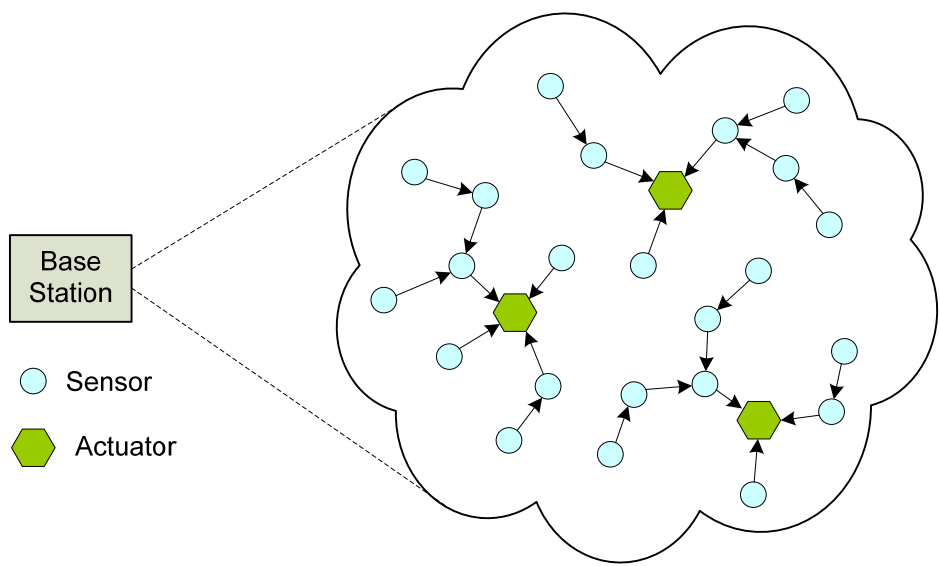

**Figure 1.** A wireless sensor and actuator network.

As shown in Figure 1, a WSAN is a networked system of geographically distributed sensor and actuator nodes that are interconnected via wireless links. Both sensor and actuator nodes are normally equipped with certain data processing and wireless communication capabilities, as well as power supply. In most situations, sensor nodes are stationary whereas actuator nodes, e.g., mobile robots and unmanned aerial vehicles, are mobile. Sensors gather information about the state of physical world and transmit the collected data to actuators through single-hop or multi-hop communications. Upon receiving the required information, the actuators make the decision about how to react to this information and perform corresponding actions to change the behavior of the physical environment. The base station is principally responsible for monitoring and managing the overall network through communications with sensors and actuators.

It is not until very recently that the area of WSAN has begun to emerge, regardless of one-decade research activities in WSN. Relatively very little work has been conducted on WSAN. In particular, quality of service (QoS) management in WSANs is an area yet to be explored [4]. WSANs are application-oriented, especially when used for cyber-physical computing. Therefore, QoS has to be supported by WSANs in order to achieve end users' satisfaction with the services that the system provides. To a large extent, the performance of future cyber-physical systems will rely on the QoS support in WSANs, just like how we today rely on various services offered by Internet to communicate with one another.

This paper gives a brief overview of QoS provisioning in the context of WSANs. Some critical challenges and possible research topics are discussed. Related work is reviewed. The primary aim is to spark new research and development interests in this field.

## 2. QoS Requirements

As Moore's law continues, it is envisioned that WSANs will become pervasive in our daily lives, for example, in our homes, offices, and cars. They promise to revolutionize the way we understand and manage the physical world, just as Internet transformed how we interact with one another. Ultimately, they will be connected to the Internet in order to achieve global information sharing [8]. This technical trend is driving WSANs to provide QoS support because they have to satisfy the service requirements of various applications.

From an end user's perspective, real-world WSAN applications have their specific requirements on the QoS of the underlying network infrastructure [4]. For instance, in a fire handling system, sensors need to report the occurrence of a fire to actuators in a timely and reliable fashion; then, the actuators equipped with water sprinklers will react by a certain deadline so that the situation will not become uncontrollable. It is intuitive that different applications may have different QoS requirements. For instance, for a safety-critical control system, large delay in transmitting data from sensors to actuators and packet loss occurring during the course of transmission may not be allowed, while they may be acceptable for an air-conditioning system that maintains the temperature inside an office.

Although QoS is an overused term, there is no common or formal definition of this term. Conceptually, it can be regarded as the capability to provide assurance that the service requirements of applications can be satisfied. Depending on the type of target application, QoS in WSANs can be characterized by reliability, timeliness, robustness, availability, and security, among others. Some QoS parameters may be used to measure the degree of satisfaction of these services, such as throughput, delay, jitter, and packet loss rate. There are many other QoS parameters worth mentioning, but these four are the most fundamental [9-12].

- *Throughput* is the effective number of data flow transported within a certain period of time, also specified as bandwidth in some situations. In general, the bigger the throughput of the network, the better the performance of the system is. Those nodes that generate high-speed data streams, such as a camera sensor node used to transmit images for target tracking, often require high throughput. In order to improve the resource efficiency, furthermore, the throughput of WSAN should often be maximized.
- *Delay* is the time elapsed from the departure of a data packet from the source node to the arrival at the destination node, including queuing delay, switching delay, propagation delay, etc. Delay-sensitive applications usually require WSANs to deliver the data packets in real-time. Notice that real-time does not necessarily mean fast computation or communication [5,12]. A real-time system is unique in that it needs to execute at a speed that fulfills the timing requirements.
- *Jitter* is generally referred to as variations in delay, despite many other definitions. It is often caused by the difference in queuing delays experienced by consecutive packets.
- *Packet loss rate* is the percentage of data packets that are lost during the process of transmission. It can be used to represent the probability of packets being lost. A packet may be lost due to e.g. congestion, bit error, or bad connectivity. This parameter is closely related to the reliability of the network.

## 3. Challenges

WSANs cannot be simply regarded as WSNs due to the co-existence of sensors and actuators, as mentioned previously. In this section, some of the major features of WSANs that challenge QoS provisioning will be discussed.

*3.1. Resource Constraints*

As in WSNs, sensor nodes are usually low-cost, low-power, small devices that are equipped with only limited data processing capability, transmission rate, battery energy, and memory. For example, the MICAz mote from Crossbow is based on the Atmel ATmega128L 8-bit microcontroller that provides only up to 8 MHz clock frequency, 128-KB flash program memory and 4-KB EEPROM; the transmit data rate is limited to 250 Kbps. Due to the limitation on transmission power, the available bandwidth and the radio range of the wireless channel are often limited. In particular, energy conservation is critically important for extending the lifetime of the network, because it is often infeasible or undesirable to recharge or replace the batteries attached to sensor nodes once they are deployed. Actuator nodes typically have stronger computation and communication capabilities and more energy budget relative to sensors. Resource constraints apply to both sensors and actuators, notwithstanding.

In the presence of resource constraints, the network QoS may suffer from the unavailability of computing and/or communication resources. For instance, a number of nodes that want to transmit messages over the same WSAN have to compete for the limited bandwidth that the network is able to provide. As a consequence, some data transmissions will possibly experience large delays, resulting in low level of QoS. Due to the limited memory size, data packets may be dropped before the nodes successfully send them to the destination. Therefore, it is of critical importance to use the available resources in WSANs in a very efficient way.

*3.2. Platform Heterogeneity*

Sensors and actuators do not share the same level of resource constraints, as mentioned above. Possibly designed using different technologies and with different goals, they are different from each other in many aspects such as computing/communication capabilities, functionality, and number. In a large-scale system of systems, the hardware and networking technologies used in the underlying WSANs may differ from one subsystem to another. This is true because of the lack of relevant standards dedicated to WSANs and hence commercially available products often have disparate features. This platform heterogeneity makes it very difficult to make full use of the resources available in the integrated system. Consequently, resource efficiency cannot be maximized in many situations. In addition, the platform heterogeneity also makes it challenging to achieve real-time and reliable communication between different nodes.

*3.3. Dynamic Network Topology*

Unlike WSNs where (sensor) nodes are typically stationary, the actuators in WSANs may be mobile. In fact, node mobility is an intrinsic nature of many applications such as, among others, intelligent transportation, assisted living, urban warfare, planetary exploration, and animal control. During runtime, new sensor/actuator nodes may be added; the state of a node is possibly changed to or from sleeping mode by the employed power management mechanism; some nodes may even die due to exhausted battery energy. All of these factors may potentially cause the network topologies of WSANs to change dynamically.

Dealing with the inherent dynamics of WSANs requires QoS mechanisms to work in dynamic and even unpredictable environments. In this context, QoS adaptation becomes necessary; that is, WSANs must be adaptive and flexible at runtime with respect to changes in available resources. For example, when an intermediate node dies, the network should still be able to guarantee real-time and reliable communication by exploiting appropriate protocols and algorithms.

*3.4. Mixed Traffic*

Diverse applications may need to share the same WSAN, inducing both periodic and aperiodic data. This feature will become increasingly evident as the scale of WSANs grows. Some sensors may be used to create the measurements of certain physical variables in a periodic manner for the purpose of monitoring and/or control. Meanwhile, some others may be deployed to detect critical events. For instance, in a smart home, some sensors are used to sense the temperature and lighting, while some others are responsible for reporting events like the entering or leaving of a person. Furthermore, disparate sensors for different kinds of physical variables, e.g., temperature, humidity, location, and speed, generate traffic flows with different characteristics (e.g. message size and sampling rate). This feature of WSANs necessitates the support of service differentiation in QoS management.

## 4. Open Issues

Over the years, in order to meet the requirements of diverse applications on network QoS, significant effort has been made to provide end-to-end QoS support using various algorithms and mechanisms at different network protocol layers. Particularly, Internet QoS has been a focus of enormous research and development activities [9]. Due to the many distinctive characteristics of WSANs, however, existing QoS mechanisms may not be applicable to WSANs [10,11]. To achieve QoS support in WSAN, the above challenges have to be addressed. In this section, several open research topics of interest will be identified.

*4.1. Service-Oriented Architecture*

The concept of service-oriented architecture (SOA) [6,13-15] is by no means new and has been widely used in for example the web services domain. However, many of its elegant potentials have not ever been explored in WSANs, though SOA will undoubtedly have a major impact in many branches of technology [16]. SOA is an architectural style encompassing a set of services for building complex systems of systems. It can be regarded as a model in which a system is decomposed into smaller, distinct units that are able to provide certain functionality. As an architectural evolution and a paradigm shift in systems integration, SOA enables rapid, cost-effective composition of interoperable, scalable systems based on reusable services exposed by these systems. This is particularly useful for QoS provisioning in WSANs that are integrated into large-scale cyber-physical systems in which multiple applications run on diverse technologies and platforms.

Identifying and specifying services are crucial for exploiting SOA in WSANs. A large number of questions need to be answered in this respect. For example, how many categories of services should be classified in the context of WSAN? What are the functionality, interface, and properties of each service? What are its quality levels relevant to performance requirements? In particular, how to deal with the difference between sensors and actuators when specifying services?

*4.2. QoS-Aware Communication Protocols*

In order to efficiently support QoS in WSANs, communication protocols need to be designed with in mind the platform heterogeneity, specifically the heterogeneity between sensors and actuators that are involved in the communication. For this reason, state-of-the-art QoS-aware MAC, routing, and transport protocols devoted to WSNs may not be suitable for WSANs.

As an essential component of QoS, service differentiation should be supported by communication protocols. As mentioned above, WSANs may be used in cyber-physical systems encompassing diverse applications, which may differ significantly in terms of QoS requirements. Obviously, the best-effort service offered by current wireless networking technologies such as Zigbee and Bluetooth cannot provide different QoS to different applications. Therefore, the communication protocols for WSANs should be designed to perceive the service requirement of each type of traffic so that it can be guaranteed a specific service level. From a practice perspective, the best-effort service is likely to be the standard for the foreseeable future [9,10]. It is therefore necessary for all new QoS mechanisms to be layered on top of the existing networks.

Cross-layer design has proved to be effective in optimizing the network performance and hence may be incorporated in the development of QoS-aware communication protocols for WSANs. Much work can be conducted in this line. For example, the prioritization of traffic at lower layers might be associated with the application performance at the application layer.

*4.3. Resource Self-Management*

Resource management is of paramount importance for QoS provisioning because the corresponding resource budgets need to be guaranteed in order to achieve certain QoS levels. This is particularly true for WSANs where computing, communication and energy resources are inherently limited. Generally speaking, a higher level of QoS corresponds to a need of more resources, e.g. CPU time, memory size, bandwidth and/or energy. Resource management in WSANs is challenging, because of the ever-increasing complexity of WSANs, highly dynamic feature of WSANs, and changing and unpredictable environments in which WSANs operate.

To overcome these challenges, self-management technologies [17,18] are needed. This implies that the system will address resource management issues in an autonomous manner. With respect to changes in resource availability, resource manager will automatically adapt resource usage in a way that the resulting overall QoS is optimized. This has to be performed in an efficient way. Since the resources are limited, the overhead of resource management should be minimized. In order to maintain scalability, distributed mechanisms have to be explored in this context.

A promising way to go is to exploit feedback scheduling [4,19-21]. Taking advantage of well-established control theory and technology, feedback scheduling offers a promising approach to flexible resource management in dynamic and unpredictable environments. Previous work has showed that feedback scheduling is capable of handling uncertainties in resource availability through automatically adapting to dynamic changes. It is anticipated that this technology can be used in WSANs to realize *resource self-management* and to provide QoS guarantees. The predictability of the system can be enhanced thanks to the use of control theory. Nevertheless, how to map resource management to control problems is still subject of future research.

*4.4. QoS-Aware Power Management*

Energy conservation is a major concern in both WSNs and WSANs. The lifetime of untethered sensor/actuator nodes is tightly restricted by the available battery energy. Since wireless communication is much more energy-expensive than sensing and computation, the transmission power of nodes has to be properly managed in a way that the energy consumption is minimized in order to prolong the lifetime of the whole network. Due to the increasingly heavy computational burdens, a significant amount of energy will be consumed by the computations in actuator nodes. Therefore, the CPU energy consumption of actuators should also be minimized, e.g. by exploiting the dynamic voltage scaling technology.

However, minimizing energy consumption and maximizing QoS are in most cases two conflicting requirements. For instance, reliability can be improved by increasing the number of maximum allowable retransmissions or using higher transmission power levels; however, more energy will be expended in both cases. Therefore, tradeoffs must be made between energy conservation and QoS optimization. The problem then becomes how to make these tradeoffs at runtime. Is it possible to find an integrated performance metric that accounts for both energy efficiency and QoS, and then optimize it, either online or offline?

Depending on the network topology and the QoS requirements, the power management mechanisms for actuator nodes may be different from those used in sensor nodes. Thus the QoS can be maximized through exploiting the different capabilities of sensors and actuators. In like manner, different transmission power levels may be assigned to the same node with respect to different types of traffic. In-network computation can be exploited to reduce the energy consumption of both sensor and actuator nodes since it reduces traffic load at the cost of slightly increased computation in each involved node. Still, the inherently non-deterministic and open nature of wireless channels poses great challenges for QoS-aware power management.

*4.5. Supporting Tools*

The fundamental role of WSAN is to connect the cyber space and the physical world. Cyberspaces are by nature discrete-time systems, whereas the physical world is composed of continuous-time systems. This hybrid feature of the integrated system challenges the development of simulation and design tools that can be used to evaluate the performance of QoS mechanisms for WSANs. An interesting question is whether or not it is technically feasible to develop such a tool based on a service-oriented architecture. If so, programming technologies for implementing various services need to be developed. In addition, benchmark testbeds and prototypes also deserve extensive research and development effort. Using these supporting tools, guidelines can be further developed that help implement new protocols, mechanisms, and algorithms for QoS management in practice.

## 5. Recent Progress

Service-oriented approaches have been used in building WSANs. In [6], Rezgui and Eltoweissy explored the potential of SOA in building open, efficient, interoperable, scalable, and application-aware WSANs. A prototype service-oriented WSAN was developed on top of TinyOS. King *et al* [22] developed a service-oriented WSAN platform called Atlas, which enables self-integrative, programmable pervasive spaces. Kushwaha *et al* [14] developed a programming framework called OASiS that provides abstractions for object-centric, ambient-aware, service-oriented sensor network applications. OASiS decomposes specified application behavior and generates the appropriate node-level code for deployment onto sensor networks. It enables the development of sensor network applications without having to deal with the complexity and unpredictability of low-level system and network issues. Chu and Buyya [15] presented a reusable, scalable, extensible, and interoperable service-oriented sensor web architecture. The architecture conforms to the sensor web enablement standard defined by the OpenGIS consortium (OGC), integrates sensor web with grid computing, and provides middleware support for sensor webs. Golatowski *et al* [23] proposed a service-oriented software architecture for mobile sensor networks. An adaptive middleware is employed in the architecture that encompasses mechanisms for cooperative data mining, self-organization, networking, and energy optimization to build higher-level service structures.

Some efforts have been made on communication protocols that provide QoS support in WSANs. Real-time, reliable communication has been addressed in e.g. [24-32]. Ngai *et al* [24] designed a real-time communication framework that supports event detection, reporting, and actuator coordination. Shah *et al* [25] proposed a real-time coordination and routing framework that addresses the coordination of sensors and actuators and respects the delay bound for routing in a distributed manner. Melodia *et al* [26] presented a distributed protocol for sensor-actor coordination that includes an adaptive mechanism to trade off energy consumption for delay when the data transmission is subject to real-time constraints. Hu *et al* [27] developed an anycast communication paradigm that can reduce both end-to-end latency and energy consumption. A simple yet effective wireless communication model has been employed in [28] that realizes real-time actuation in autonomous animal control. Boukerche *et al* [29] proposed a routing protocol with service differentiation for WASNs, which provides low latency and reliable delivery in the presence of failures. Morita *et al* [30] presented a redundant data transmission protocol that can significantly enhance the reliability of data transmission over lossy WSANs. A general reliability-centric framework for event reporting has been presented in [31]. A low-complexity reliable transmission scheme has been developed in [32], which is based on local wireless path repair and hop-to-hop retransmissions.

Adaptive sampling approaches have been exploited for dynamic management of resources in WSNs, e.g. [33,34]. However, these approaches don't take into account the co-existence of sensors and actuators. Few solutions have been devised for resource self-management that facilitates QoS-enabled autonomic WSAN. In [4,35], Xia *et al* applied feedback control technologies to dynamic bandwidth allocation in the context of WSAN, which take advantage of the idea of feedback scheduling. The flexibility and autonomy of the system is enhanced through deadline miss ratio control, leading to improved QoS in terms of reliability.

While QoS-aware power management in general WSNs has been extensively studied, e.g. [36, 37], there is relatively little work dedicated to QoS-aware power management in WSANs, particularly for actuator nodes. Rozell and Johnson [38] developed an optimal method for power scheduling in WSAN that achieves a desired actuation fidelity. Sanchez *et al* [39] proposed an energy-efficient multicast routing protocol that was specially designed to minimize the total energy used by the multicast tree. Zhou *et al* [40] developed a data transport protocol that reduces the energy consumption associated with data transmission while meeting the QoS requirements in timeliness domain. Durresi *et al* [41] proposed a delay-energy aware routing protocol that enables a flexible range of tradeoffs between the packet delay and the energy use. A power-aware many-to-many routing scheme has been proposed in [42]. A low-energy and delay-sensitive TDMA based MAC protocol has been presented in [43].

In line of supporting tools development, an interesting attempt is TrueTime [44], a Matlab/Simulink-based simulator developed at Lund University. TrueTime facilitates co-simulation of controller task execution in real-time kernels, network transmissions, and continuous plant dynamics. It was designed primarily for simulating networked and embedded control systems, and has supported sensor network applications. Another notable attempt is the Agent/Plant [45] module developed at Case Western Reserve University. The module extends NS-2 to interface network dynamics with physical behaviors, making it possible to simulate physical systems that are attached to a network.

## 6. Conclusion

WSAN is an area still in its infancy, despite some recent progress. It is anticipated that WSANs will evolve rapidly and become pervasive in the near future, much in the same way as the Internet came to the desktop before. Lessons should be taken from Internet that WSANs have to be designed with QoS support in mind. This paper has discussed the requirements and challenges for supporting QoS in WSANs. Some interesting open research topics have been identified, though the spectrum of research in this field can be much broader. The challenges are formidable and extensive research from multiple disciplines is needed before QoS-enabled WSANs become reality.

**Acknowledgements**

The author would like to thank Yu-Chu Tian, Guosong Tian, and Li Gui at Queensland University of Technology, Australia, for the many insights he have gained in working with them. This work is supported in part by Australian Research Council (ARC) under the Discovery Projects grant DP0559111 and China Postdoctoral Science Foundation under grant 20070420232.

**References and Notes**

1. Borriello, G.; Farkas, K.I.; Reynolds, F.; Zhao, F. Building A Sensor-Rich World. *IEEE Pervasive Computing* **2007**, *6*(2), 16-18.
2. Chong, C.-Y.; Kumar, S.P. Sensor networks: evolution, opportunities, and challenges. *Proceedings of the IEEE* **2003**, *91*(8), 1247-1256.
3. A wireless sensor networks bibliography. http://ceng.usc.edu/~anrg/SensorNetBib.html.
4. Xia, F.; Zhao, W.H.; Sun, Y.X.; Tian, Y.C. Fuzzy Logic Control Based QoS Management in Wireless Sensor/Actuator Networks. *Sensors* **2007**, *7*(12), 3179-3191.
5. Xia, F.; Tian, Y.C; Li, Y.J.; Sun, Y.X. Wireless Sensor/Actuator Network Design for Mobile Control Applications. *Sensors* **2007**, *7*(10), 2157-2173.
6. Rezgui, A.; Eltoweissy, M. Service-Oriented Sensor-Actuator Networks. *IEEE Communications Magazine* **2007**, *45*(12), 92-100.
7. Akyildiz, I. F.; Kasimoglu, I. H. Wireless sensor and actor networks: research challenges. *Ad Hoc Networks* **2004**, *2*(4), 351-367.
8. NSF Workshop on Cyber-Physical Systems, http://varma.ece.cmu.edu/cps/, Oct. 2006
9. El-Gendy, M. A.; Bose, A.; Shin, K. G. Evolution of the Internet QoS and support for soft real-time applications. *Proceedings of the IEEE* **2003**, *91*(7), 1086-1104.
10. Chen, D.; Varshney, P. K. QoS Support in Wireless Sensor Networks: A Survey. In Proc. of the Int. Conf. on Wireless Networks, Las Vegas, USA, June 2004.
11. Li, Y.J.; Chen, C.S.; Song, Y.-Q.; Wang, Z. Real-time QoS support in wireless sensor networks: a survey. In Proc of 7th IFAC Int Conf on Fieldbuses & Networks in Industrial & Embedded Systems (FeT'07), Toulouse, France, Nov. 2007.
12. Bouyssounouse, B.; Sifakis, J. (eds.) Embedded Systems Design: The ARTIST Roadmap for Research and Development. Lecture Notes in Computer Science 3436, Springer-Verlag, 2005.
13. Papazoglou, M.P. Service-Oriented Computing: Concepts, Characteristics and Directions. In Proc. of the Fourth Int. Conf. on Web Information Systems Engineering, Dec 2003; pp. 3-12.

14. Kushwaha, M.; Amundson, I.; Koutsoukos, X.; Neema, S.; Sztipanovits, J. OASiS: A Programming Framework for Service-Oriented Sensor Networks. In Proc. 2nd Int. Conf. on Communication Systems Software and Middleware, Bangalore, India, Jan. 2007; pp.1-8.
15. Chu, X.; Buyya, R. Service Oriented Sensor Web. In: Mahalik, N. P. (ed), Sensor Network and Configuration: Fundamentals, Standards, Platforms, and Applications. Springer-Verlag, ISBN: 978-3-540-37364-3, Germany, Jan. 2007; pp.51-74.
16. MORE Project Deliverable D2.1: Architecture and Services, http://www.ist-more.org/.
17. Ganek, A. G.; Corbi, T. A. The dawning of the autonomic computing era. *IBM Systems Journal* **2003**, *42*(1), 5-18.
18. Herrmann, K.; Muhl, G.; Geihs, K. Self management: the solution to complexity or just another problem. *IEEE Distributed Systems Online* **2005**, *6*(1), 1-17.
19. Xia, F.; Tian, G.S.; Sun, Y.X. Feedback Scheduling: An Event-Driven Paradigm. *ACM SIGPLAN Notices* **2007**, *42* (12), 7-14.
20. Xia, F. Feedback scheduling of real-time control systems with resource constraints, PhD thesis, Zhejiang University, 2006.
21. Arzen, K.-E.; Robertsson, A.; Henriksson, D.; Johansson, M.; Hjalmarsson, H.; Johansson, K.H. Conclusions of the ARTIST2 Roadmap on Control of Computing Systems. *ACM SIGBED Review* **2006**, *3*(3), 11-20.
22. King, J.; Bose, R.; Yang, H.; Pickles, S.; Helal, A. Atlas: a service-oriented sensor platform. In Proc. of 31st IEEE Conf. on Local Computer Networks, Nov 2006; pp. 630-638.
23. Golatowski, F.; Blumenthal, J.; Handy, M.; Haase, M.; Burchardt, H.; Timmermann, D. Service-Oriented Software Architecture for Sensor Networks. In Proc. Int. Workshop on Mobile Computing (IMC'03), Rockstock, Germany, June 2003; pp. 93-98.
24. Ngai, E. C.H.; Lyu, M.R.; Liu, J. A Real-Time Communication Framework for Wireless Sensor-Actuator Networks. In Proc. IEEE Aerospace Conf., Big Sky, Montana, U.S.A., March 2006.
25. Shah, G.A.; Bozyigit, M.; Akan, O.B.; Baykal, B. Real-Time Coordination and Routing in Wireless Sensor and Actor Networks. In Proc. 6th Int. Conf. on Next Generation Teletraffic and Wired/Wireless Advanced Networking (NEW2AN), Lecture Notes in Computer Science, 2006; Vol. 4003, pp. 365-383.
26. Melodia, T.; Pompili, D.; Gungor, V. C.; Akyildiz, I. F. Communication and Coordination in Wireless Sensor and Actor Networks. *IEEE Transactions on Mobile Computing* **2007**, *6*(10), 1116-1129.
27. Hu, W.; Bulusu, N.; Jha, S. A Communication Paradigm for Hybrid Sensor/Actuator Networks. *International Journal of Wireless Information Networks* **2005**, *12*(1), 47-59.
28. Wark, T.; Crossman, C.; Hu, W.; Guo, Y.; Valencia, P.; Sikka, P.; Corke, P.I.; Lee, C.; Henshall, J.; Prayaga, K.; O'Grady, J.; Reed, M.; Fisher, A. The design and evaluation of a mobile sensor/actuator network for autonomous animal control. In Proc. Int. Conf. on Information Processing in Sensor Networks (IPSN), 2007; pp. 206-215.
29. Boukerche, A.; Araujo, R.B.; Villas, L. A Wireless Actor and Sensor Networks QoS-Aware Routing Protocol for the Emergency Preparedness Class of Applications. In Proc. 31st IEEE Conf.on Local Computer Networks, Tampa, FL, 2006; pp. 832-839.
30. Morita, K.; Ozaki, K.; Hayashibara, N.; Enokido, T.; Takizawa, M. Evaluation of Reliable Data Transmission Protocol in Wireless Sensor-Actuator Network. In Proc. 21st Int. Conf. on Advanced Information Networking and Applications Workshops, May 2007; Vol. 2, pp. 713-718.

31. Ngai, E. C.H.; Zhou, Y.; Lyu, M.R.; Liu, J. Reliable Reporting of Delay-Sensitive Events in Wireless Sensor-Actuator Networks. In Proc. of the 3rd IEEE Int. Conf. on Mobile Ad-Hoc and Sensor Systems (MASS'06), Vancouver, Canada, Oct. 2006.
32. Hu, F.; Cao, X.; Kumar, S.; Sankar, K. Trustworthiness in wireless sensor and actuator networks: towards low-complexity reliability and security. In Proc. IEEE Global Telecommunications Conference (GLOBECOM), Vol.3, Dec. 2005.
33. Gedik, B.; Liu, L.; Yu, P.S. ASAP: An Adaptive Sampling Approach to Data Collection in Sensor Networks. *IEEE Transactions on Parallel and Distributed Systems* **2007**, *18*(12), 1766-1783.
34. Liu, X.; Wang, Q.; He, W.; Caccamo, M.; Sha, L. Optimal Real-Time Sampling Rate Assignment for Wireless Sensor Networks. *ACM Transactions on Sensor Networks* **2006**, *2*(2), 263-295.
35. Xia, F.; Zhao, W.H. Flexible Time-Triggered Sampling in Smart Sensor-Based Wireless Control Systems. *Sensors* **2007**, *7*(11), 2548-2564.
36. Lin, S.; Zhang, J.; Zhou, G.; Gu, L.; He, T.; Stankovic, J.A. ATPC: Adaptive Transmission Power Control for Wireless Sensor Networks. In Proc. of SenSys'06, Boulder, Colorado, USA, Nov. 2006.
37. Vidhyapriya R.; Vanathi, P.T. Conserving energy in wireless sensor networks. *IEEE Potentials* **2007**, *26*(5), 37-42.
38. Rozell, C.J.; Johnson, D.H. Power scheduling for wireless sensor and actuator networks. In Proc. of IPSN, Cambridge, MA, April 2007.
39. Sanchez, J.A.; Ruiz, P.M.; Stojmenovic, I. Energy-efficient geographic multicast routing for Sensor and Actuator Networks. *Computer Communications*, **2007**, *30*, 2519-2531.
40. Zhou, Y.; Ngai, E. C.-H.; Lyu, M.R.; Liu, J. POWER-SPEED: A Power-Controlled Real-Time Data Transport Protocol for Wireless Sensor-Actuator Networks. In Proc. IEEE Wireless Communications and Networking Conf (WCNC'07), Hong Kong, China, March 2007.
41. Durresi, A.; Paruchuri, V.; Barolli, L. Delay-Energy Aware Routing Protocol for Sensor and Actor Networks. In Proc. of 11th Int. Conf. on Parallel and Distributed Systems, Fuduoka, Japan, July 2005; pp. 292-298.
42. Cayirci, E.; Coplu, T.; Emiroglu, O. Power aware many to many routing in wireless sensor and actuator networks. In Proc. of the Second European Workshop on Wireless Sensor Networks, 2005; pp. 236-245.
43. Munir, M.F.; Filali, F. Low energy, adaptive and distributed MAC protocol for wireless sensor-actuator networks. In Proc. of 18th IEEE Annual Int. Symposium on Personal Indoor and Mobile Radio Communications, Athens, Greece, Sept. 2007.
44. http://www.control.lth.se/truetime/.
45. http://vorlon.case.edu/~vxl11/NetBots/.